\documentclass[showpacs,epsf,twocolumn]{revtex4-1}
\usepackage{float}
\usepackage{color,graphicx}
\graphicspath{ {images/} }
\usepackage{hyperref}
\begin{document}

\title{Magnetotransport properties and giant anomalous Hall angle in half-Heusler compound TbPtBi}

\author{Ratnadwip Singha, Shubhankar Roy, Arnab Pariari, Biswarup Satpati and Prabhat Mandal}
\email{prabhat.mandal@saha.ac.in}
\affiliation{Saha Institute of Nuclear Physics, HBNI, 1/AF Bidhannagar, Calcutta 700 064, India}
\date{\today}

\begin{abstract}
Magnetic lanthanide half-Heuslers ($R$PtBi; $R$ being the lanthanide) represent an attractive subgroup of the Heusler family and have been identified as ideal candidates for time reversal symmetry breaking topological Weyl semimetals. In this paper, we present the detailed analysis of the magnetotransport properties of frustrated antiferromagnet TbPtBi. This material shows large, non-saturating magnetoresistance (MR) with unusual magnetic field dependence. The MR of TbPtBi is significantly anisotropic with respect to the magnetic field, applied along different crystallographic directions and indicates the anisotropic nature of the Fermi surface. The chiral anomaly induced negative longitudinal magnetoresistance confirms the presence of Weyl fermions. At low temperature, Berry phase driven large anomalous Hall conductivity has been observed. The calculated anomalous Hall angle is the largest reported so far.
\end{abstract}

\maketitle

\section{Introduction}

The family of Heusler compounds consists of more than 1000 ternary metallic or semiconducting materials with wide range of multifunctional properties \cite{Graf}. Depending on the chemical composition, they show different physical phenomena varying from superconductivity \cite{Klimczuk,Winterlik,Xu}, half-metallic ferromagnetism \cite{Groot}, heavy fermionic behavior \cite{Canfield,Fisk}, compensated ferrimagnetism \cite{Wurmehl} to more recently topological insulating/semimetallic states \cite{Chadov,Nakajima}. Hence, different members of this family have been extensively investigated since past few decades and often found to be ideal candidates for spintronics, magnetic memory and thermoelectric applications as well as to probe exotic phases such as topological superconductors and magnetic Weyl semimetals. Moreover, the large number of isostructural systems provide immense tunability of the desired electronic state by changing the hybridization strength or the spin-orbit coupling. Half-Heusler compounds are a subgroup of this family with general chemical formula \textit{XYZ}, where $X$ and $Y$ are either transition metal or rare earth element and $Z$ is a main group element. Several half-Heusler materials have been theoretically proposed \cite{Chadov,Xiao} and in some cases experimentally verified \cite{Liu,Logan} to be topological insulators. On the other hand, the signatures of Weyl fermions have been reported in very few members \cite{Hirschberger,Shekhar}. Weyl nodes act as monopole source/sink of Berry curvature and appear at the crossing point of linearly dispersing energy bands in absence of inversion (IS) or time reversal symmetry (TRS). While majority of the Weyl semimetals, discovered so far, lack IS \cite{Lv,Xu2,Xu3}, materials with broken TRS are very rare \cite{Borisenko,Wang}. Interestingly, in magnetic lanthanide half-Heusler compounds ($R$PtBi; $R$ is the lanthanide element), Weyl node emerges in absence of TRS due to the combination of large exchange field from the 4$f$ electrons and Zeeman splitting induced by the external magnetic field \cite{Shekhar}. These compounds are expected to show an intrinsic anomalous Hall conductivity (AHC) originating from the Berry curvature of the electronic band structure \cite{Burkov}. In fact, very large anomalous Hall angle (AHA) has already been reported in GdPtBi \cite{Shekhar,Suzuki}. Furthermore, GdPtBi also shows several novel phenomena such as chiral anomaly induced longitudinal magnetoresistance (LMR) \cite{Hirschberger} and planar Hall effect \cite{Kumar} due to its topology protected electronic state. It is yet to be established whether giant AHA is a generic feature of this series of compounds, as other members of this family have not been explored in detail. Substituting gadolinium by other rare earth elements should change the strength of magnetic interaction as well as electronic state and hence, will effectively tune these unique properties. In this paper, we have studied the magnetotransport properties of single crystalline TbPtBi. This material shows an antiferromagnetic ordering below 3.4 K. A large and anisotropic transverse MR has been observed with unusual magnetic field dependence. The Adler-Bell-Jackiw (ABJ) chiral anomaly in longitudinal MR, confirms the presence of Weyl fermions in the system. The Hall resistivity indicates hole-type carriers along with an AHC component. The calculated AHA is largest not only among the members of the $R$PtBi family but also among other materials reported so far.\\

\section{Experimental details}

The single crystals of TbPtBi were grown in Bi-flux \cite{Canfield2}. High purity elemental Tb (Alfa Aesar 99.9\%), Pt (Alfa Aesar 99.99\%) and Bi (Alfa Aesar 99.999\%) were taken in an alumina crucible in 1:1:9 molar ratio. The crucible was sealed in a quartz tube under vacuum. The quartz tube was heated to 1000$^{\circ}$C, soaked at this temperature for 12 hrs and then slowly cooled (5$^{\circ}$C/h) to 600$^{\circ}$C. At this temperature, the excess bismuth was decanted using a centrifuge. Single crystals with preferred growth direction along [111] axis and dimensions in the range 1-2 mm were obtained. The as-grown crystals were characterized by powder x-ray diffraction (XRD) technique in a Rigaku x-ray diffractometer (TTRAX III) using Cu $K\alpha$ radiation and high-resolution transmission electron microscopy (HRTEM) in an FEI, TECNAI G$^{2}$ F30, S-TWIN microscope operating at 300 kV and equipped with a GATAN Orius SC1000B CCD camera. The elemental composition was checked by energy-dispersive x-ray spectroscopy (EDX) using the same microscope with a scanning unit and a high-angle annular dark-field scanning
(HAADF) detector from Fischione (Model 3000). The magnetic measurements were done in a 7 T SQUID-VSM (MPMS3, Quantum Design). The transport measurements were performed by four-probe technique in a 9 T physical property measurement system (Quantum Design) with sample rotator option. The electrical contacts were made using gold wires and conducting silver paste.\\

\section{Results and discussion}

As shown in Fig. 1(a), TbPtBi crystallizes in a cubic structure with space group $F\bar{4}3m$ (space group no. 216) \cite{Haase}. The unit cell consists of three interpenetrating face-centered cubic lattices of each constituent element. Along [111] direction, the spins of Tb atoms are antiferromagnetically coupled. Fig. 1(b) illustrates the XRD spectrum of the powdered single crystals. The XRD pattern has been analyzed with Rietveld structural refinement using the FULLPROF software package. The refined lattice parameters $a$=$b$=$c$=6.664(2) {\AA} are consistent with the earlier report \cite{Haase}. Within the experimental resolution, the absence of any impurity peak confirms the phase purity of the grown crystals. In Fig. 2(a), we have shown the HRTEM image of a typical single crystal, which reveals the high crystalline quality of the samples. Fig. 2(b) illustrates the selected area electron diffraction pattern created by the crystallographic planes. The obtained results from EDX spectroscopy at different randomly selected regions confirm the homogeneous distribution of the elements and almost perfect stoichiometry (Tb:Pt:Bi=1:1.1:0.9) of the grown crystals. In Fig. 2(c) we have shown one such area as a representative.

\begin{figure}
\includegraphics[width=0.5\textwidth]{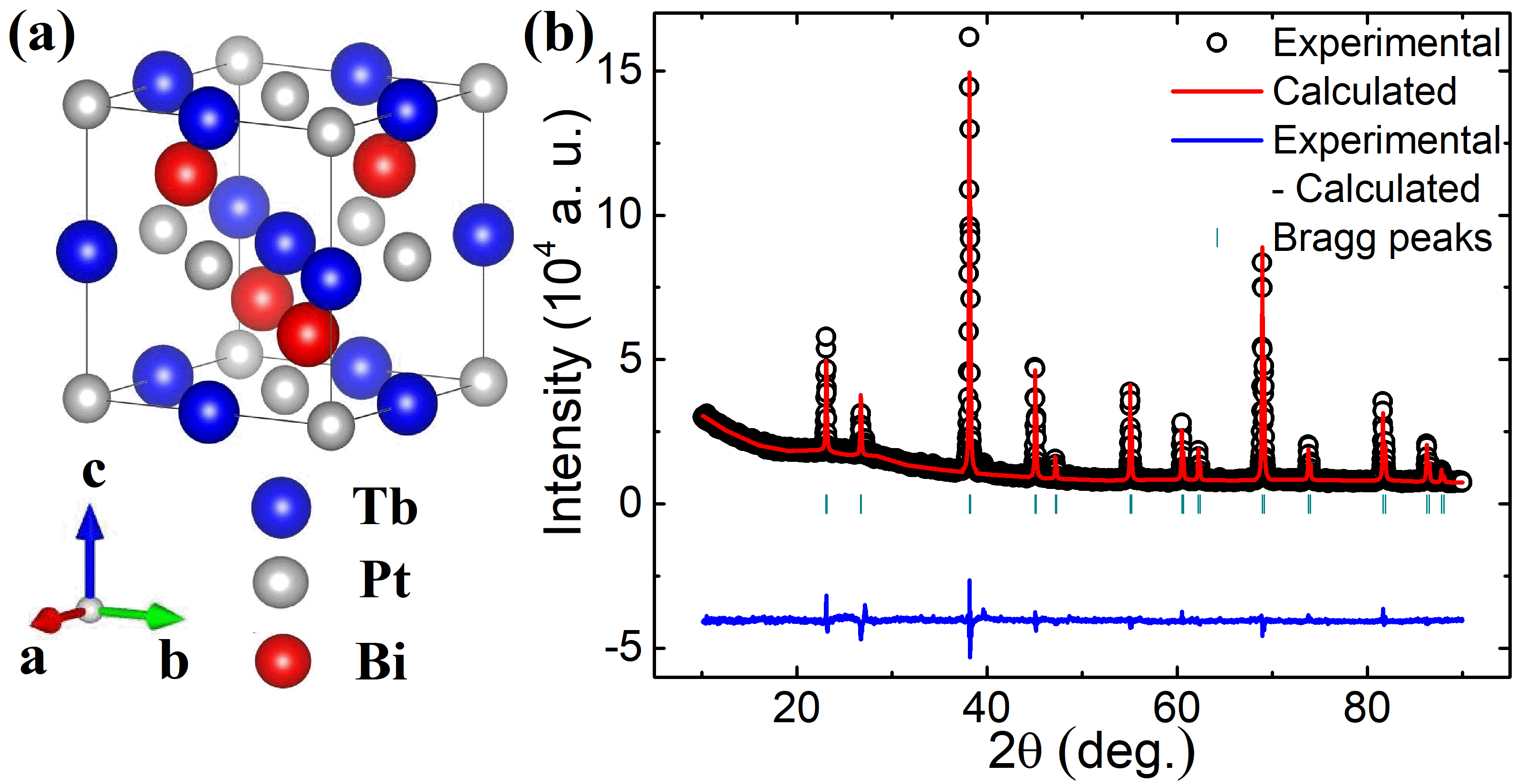}
\caption{(a) Crystal structure of TbPtBi. (b) X-ray diffraction spectra of the powdered single crystals.}
\end{figure}
\begin{figure}
\includegraphics[width=0.5\textwidth]{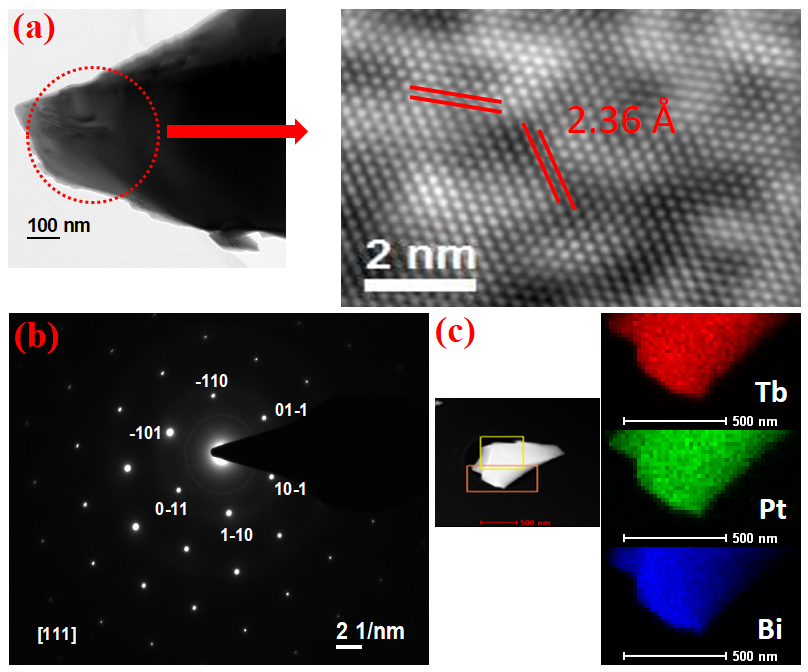}
\caption{(a) High-resolution transmission electron microscopy (HRTEM) image of the TbPtBi single crystal. (b) Selected area electron diffraction pattern obtained through HRTEM measurement. (c) Elemental mapping of a randomly selected area through energy-dispersive x-ray spectroscopy.}
\end{figure}

In Fig. 3(a), dc magnetic susceptibility ($\chi=M/B$) of TbPtBi has been plotted as a function of temperature. The experimental results clearly show an antiferromagnetic transition temperature ($T_{N}$) $\sim$3.4 K. Following the Curie-Weiss law [$\chi$=$\frac{N_{A}\mu_{eff}^{2}}{3k_{B}(T-\theta_{CW})}$], in Fig. 3(b), we have fitted the inverse susceptibility in the paramagnetic region. The x-axis intercept represents the Curie-Weiss temperature ($\theta_{CW}$) $\sim-$18 K. The calculated frustration parameter $f$=$-\theta_{N}/T_{N}$$\sim$5.3 is moderate compared to the strongly geometrically frustrated magnets ($f>$10) \cite{Ramirez,Das}. From the fitting parameters, the effective magnetic moment ($\mu_{eff}$) is found to be 9.7 $\mu_{B}$ per formula unit, which is close to the theoretical value (9.72 $\mu_{B}$) expected for Tb$^{3+}$ ion. Here, $\mu_{B}$ is the Bohr magneton.

\begin{figure}
\includegraphics[width=0.4\textwidth]{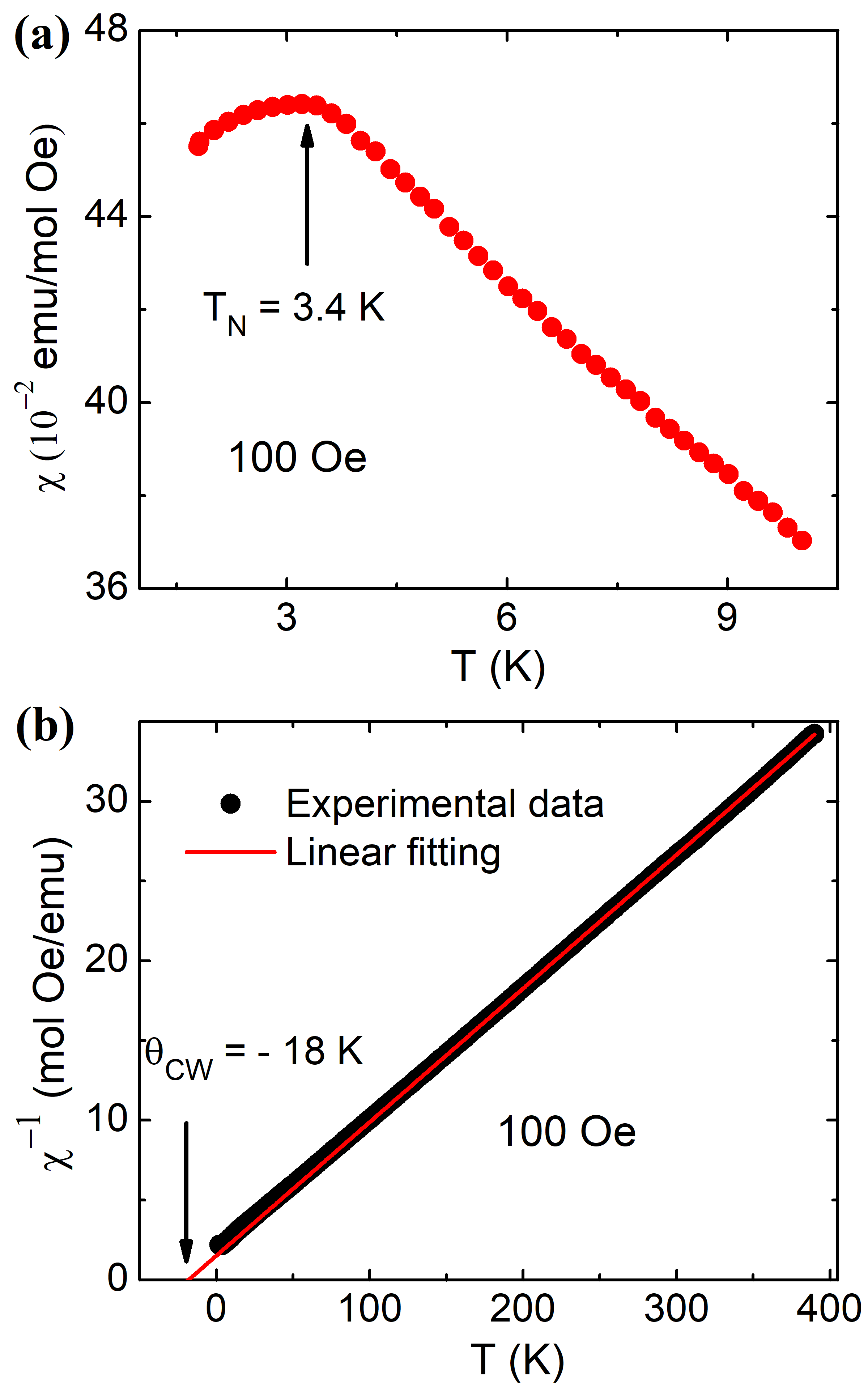}
\caption{(a) Temperature dependence of dc magnetic susceptibility of TbPtBi. The arrow shows the antiferromagnetic transition temperature ($T_{N}$). (b) Fitting of the inverse susceptibility using Curie-Weiss law. The x-axis intercept gives the Curie-Weiss temperature ($\theta_{CW}$).}
\end{figure}

\begin{figure}
\includegraphics[width=0.4\textwidth]{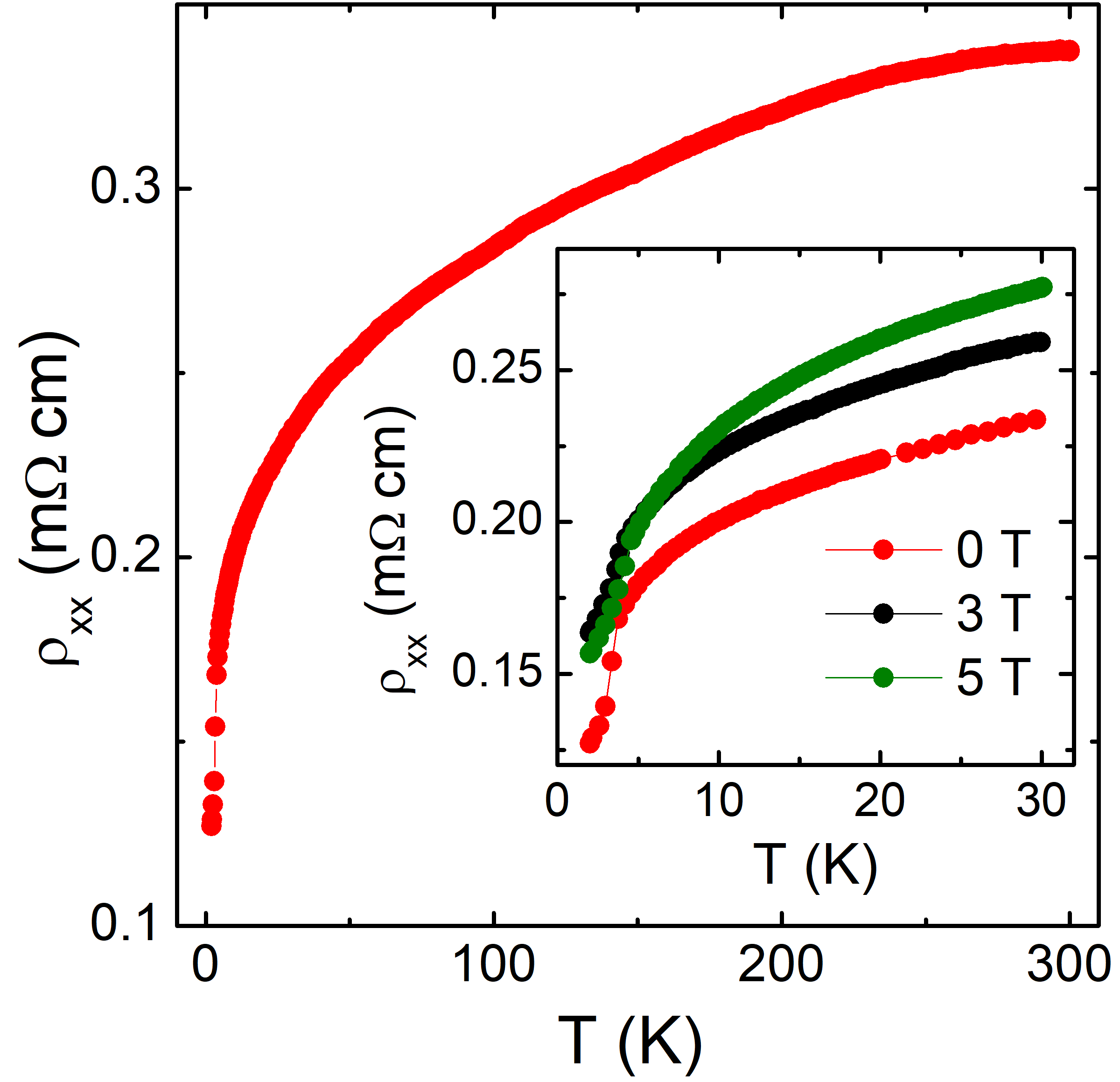}
\caption{(a) Temperature dependence of the zero-field resistivity in TbPtBi. Inset shows the resistivity for different applied magnetic field strengths.}
\end{figure}

The zero-field resistivity ($\rho_{xx}$) for TbPtBi exhibits metallic character [Fig. 4], i.e., it decreases continuously from room temperature. In an earlier study, several members of the $R$PtBi family including GdPtBi, have been reported to show semiconductor-like temperature dependence \cite{Canfield}. On the other hand, few compounds in this series such as DyPtBi and YbPtBi, do possess metallic character \cite{Canfield}. From a recent report, we note that some single crystals of GdPtBi with higher carrier density show metallic behavior \cite{Hirschberger}. In addition, the resistivity is one to two orders of magnitude smaller than the earlier report. The carrier density in TbPtBi ($\sim$10$^{19}$ cm$^{-3}$, to be discussed later) is at least one order of magnitude higher than that in GdPtBi \cite{Hirschberger} and likely to be the origin for the metallicity in this compound. As shown in the inset, just below $T_{N}$, resistivity for TbPtBi exhibits a sharp decrease. This behavior may be attributed to the suppression of spin-disorder scattering. To confirm the absence of any superconducting transition, as reported in several half-Heusler compounds \cite{Nakajima}, we have applied a magnetic field perpendicular to the current. The transition temperature remains unchanged up to the highest applied magnetic field. It is evident from the Fig. 4 inset, the magnetic field dependence of resistivity is non-monotonous in nature.

In Fig. 5(a), we have plotted $\rho_{xx}$ as a function of magnetic field at different temperatures for transverse current and magnetic field configuration. The magnetic field is applied along [111] direction and current is along [1$\bar{1}$0]. At 2 K, the resistivity shows a broad peak at $\sim$3.2 T. With increasing temperature the peak shifts towards higher field and eventually disappears above 30 K. Interestingly, the overall nature of $\rho_{xx}(B)$ curves is also sensitive to temperature. As shown in Fig. 5(b), the calculated magnetoresistance (MR) [$=\frac{\rho_{xx}(B)-\rho_{xx}(0)}{\rho_{xx}(0)}\times$100\%] reaches a maximum value $\sim$58 \% at 2 K and 9 T. The MR decreases with increasing temperature, but remains quite large (30 \%) up to 300 K. To probe any possible anisotropy in the material, we have measured the angular dependence of the transverse MR. Keeping the current direction unaltered, the magnetic field has been rotated about the [1$\bar{1}$0] axis. The experimental set-up has been shown schematically in the left panel of Fig. 5(c). The obtained results trace a two-fold symmetric pattern [Fig. 5(c) right panel]. The MR becomes as high as $\sim$80 \% at 2 K and 9 T for an angle $\sim$90$^{\circ}$ (and $\sim$270$^{\circ}$). Besides, additional kinks have been observed symmetrically at $\sim$45$^{\circ}$ (and $\sim$225$^{\circ}$) and $\sim$135$^{\circ}$ (and $\sim$315$^{\circ}$), where the MR is minimum. Similar anisotropic MR has also been reported in non-magnetic LuPtBi, which originates from the anisotropic nature of the Fermi surface \cite{Xu4}.

\begin{figure}
\includegraphics[width=0.5\textwidth]{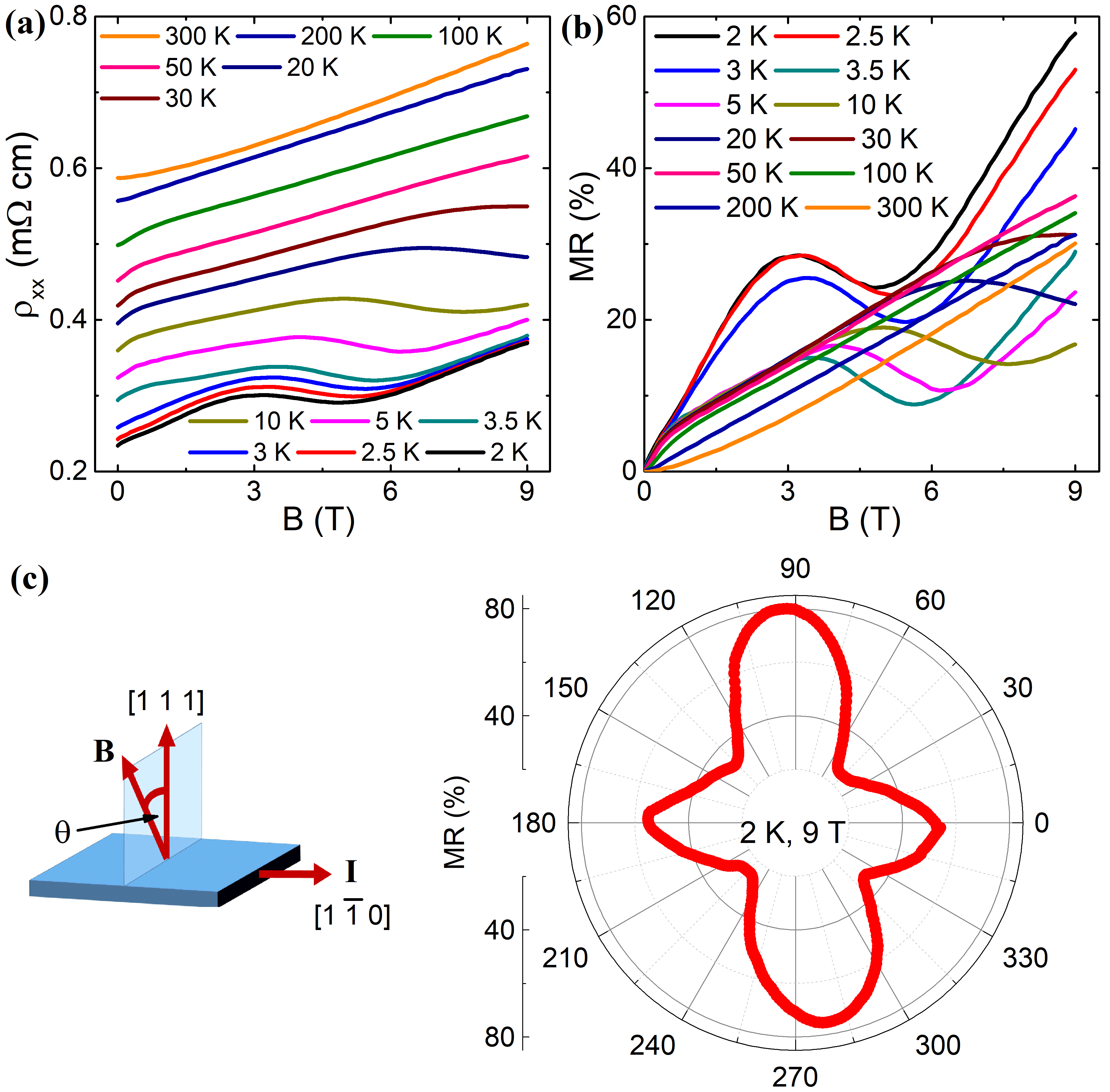}
\caption{(a) Magnetic field dependence of $\rho_{xx}$ in TbPtBi. (b) The calculated transverse magnetoresistance (TMR) at different representative temperatures. (c) Directional dependence of the TMR. The schematic shows the experimental set-up.}
\end{figure}

Identifying the presence of Weyl fermions in a material is of fundamental importance and requires sophisticated techniques such as angle resolved photo emission spectroscopy. Nevertheless, their signature can be observed in transport experiments. The ABJ chiral anomaly is a relativistic phenomenon, which appears due to the non-conservation of Weyl fermions of a definite chirality in a system \cite{Huang}. Under parallel electric and magnetic field configuration, which acts as a non-trivial gauge field, this leads to the charge pumping between two Weyl nodes of opposite chirality. The additional current flow results in conductivity enhancement and hence, is reflected as a negative MR \cite{Hirschberger,Huang,Xiong,Singha}. In Fig. 6(a), we have shown the LMR of TbPtBi, when both current and magnetic field are applied along [1$\bar{1}$0] axis. At 5 K and 9 T magnetic field, the LMR value becomes $\sim$-10 \%. The negative MR component weakens with increasing temperature and vanishes completely at 20 K. It is worth mentioning that negative MR can also originate from current jetting effect, i.e., due to the inhomogeneous current paths in the sample \cite{Hu}. To ensure the absence of current jetting, we have performed the measurements with different sets of electrical contacts and on two different single crystal samples. As illustrated in Fig. 6(b), a prominent negative LMR has also been observed in sample 2. In spin-polarized ferromagnetic materials, magnetic field suppresses the scattering of charge carriers. As a result, a negative MR is observed. In such cases, this phenomenon is independent of the angle between electric and magnetic field. However, for TbPtBi the negative MR appears only under parallel electric and magnetic field configuration. Also, the value of negative MR is expected to be maximum at the magnetic transition temperature and diminish on both side of it. However, in TbPtBi as well as GdPtBi ($T_{N}$=9.2 K) the negative MR decreases monotonically with increasing temperature \cite{Hirschberger}. For TbPtBi, the negative MR has been observed even at 15 K, i.e., at $T$/$T_{N}$ = $\sim$4, which is difficult to attribute to spin-polarized scattering. Furthermore, the chiral anomaly induced longitudinal conductivity ($\sigma^{chiral}_{xx}$) is expected to follow the semiclassical relation \cite{Huang},
\begin{equation}
\sigma^{chiral}_{xx}(B)=(1+C_{w}B^{2})(\sigma_{0}+C_{1}\sqrt{B})+(\rho_{0}+C_{2}B^{2})^{-1},
\end{equation}
where $\rho_{0}$ and $\sigma_{0}$ are zero-field resistivity and conductivity, respectively. $C_{w}$ is a temperature dependent parameter, whereas $C_{1}$ and $C_{2}$ are arbitrary constants. The term $(\sigma_{0}+C_{1}\sqrt{B})$ describes the low-field minimum in the conductivity, which appears due to weak antilocalization effect. Our experimental data are in good agreement with the theoretical prediction [Fig. 6(a) inset] and confirm the presence of Weyl fermions in TbPtBi.

\begin{figure}
\includegraphics[width=0.5\textwidth]{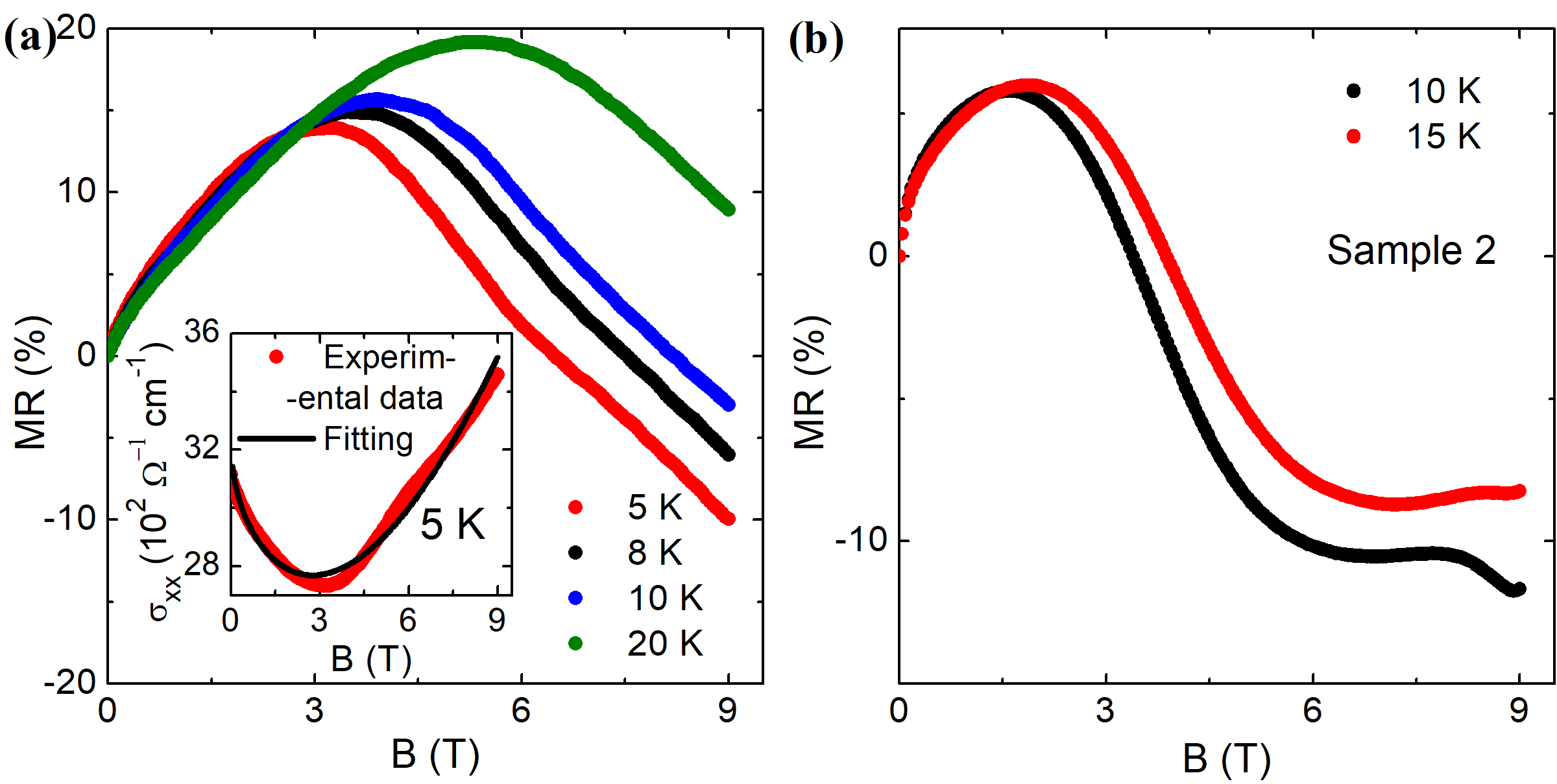}
\caption{(a) Longitudinal magnetoresistance (LMR) for TbPtBi at different temperatures. The inset shows the calculated longitudinal conductivity at 5 K. The experimental data have been fitted using Equ. (1). (b) LMR for sample 2.}
\end{figure}

To calculate the density and mobility of the charge carriers, next, we have performed the Hall resistivity ($\rho_{yx}$) measurements. In Fig. 7(a), $\rho_{yx}$ has been plotted as a function of magnetic field for several representative temperatures. Hall resistivity is positive and almost linear throughout the measured temperature range, indicating the presence of hole type majority carriers. However, a weak hump-like behavior has been observed in the low field region. With increasing temperature, this feature becomes less prominent and disappears completely above 20 K. From the linear high field region, the carrier density and mobility have been calculated to be 1.2$\times$10$^{19}$ cm$^{-3}$ and 2.2$\times$10$^{3}$ cm$^{2}$V$^{-1}$s$^{-1}$, respectively. These values are comparable to those observed in several topological semimetals \cite{Narayanan,Novak,Singha2}. To gain further insight on the nature of the charge conduction, we have calculated the total Hall angle (THA=$\sigma_{xy}/\sigma_{xx}$) for TbPtBi, where $\sigma_{xy}$ and $\sigma_{xx}$ are total Hall conductivity and longitudinal conductivity, respectively. As evident from Fig. 7(b), THA shows an anomalous magnetic field dependence. Anomalous Hall effect (AHE) is a common phenomenon in ferromagnetic systems with non-zero magnetic moment \cite{Nagaosa}. On the other hand, AHE can also be observed in antiferromagnets with non-collinear spin texture and Weyl semimetals with non vanishing Berry curvature, which acts as a fictitious magnetic field \cite{Nagaosa}. In TbPtBi, the AHE persists up to 20 K, i.e., well above the antiferromagnetic transition temperature. Hence, such behavior can not be explained from the nature of the magnetic ground state. To isolate the AHE part, we note that total Hall conductivity is simply the addition of normal ($\sigma^{N}_{xy}$) and anomalous Hall conductivity ($\sigma^{A}_{xy}$) components \cite{Suzuki}. Following the procedure for GdPtBi in earlier studies \cite{Suzuki,Shekhar}, we have used the value of $\sigma_{xy}$ for 50 K to estimate the normal Hall conductivity component in TbPtBi as AHE disappears completely at 50 K,. The calculated AHA(=$\frac{\sigma_{xy}-\sigma^{N}_{xy}}{\sigma_{xy}}$) is shown in Fig. 7(c) for different temperatures. At 2 K, the AHA becomes as large as $\sim$0.38 for a magnetic field 5.2 T. This value is largest among the members of the $R$PtBi family \cite{Suzuki,Shekhar}. In Table I, we have compared our results with other AHE compounds, which show large AHA. Note that, while the AHA is highest in TbPtBi, the AHC is lower than that for several ferromagnetic materials. With increasing temperature, the AHA decreases whereas the maximum shifts towards higher magnetic field value.\\

\begin{figure}
\includegraphics[width=0.5\textwidth]{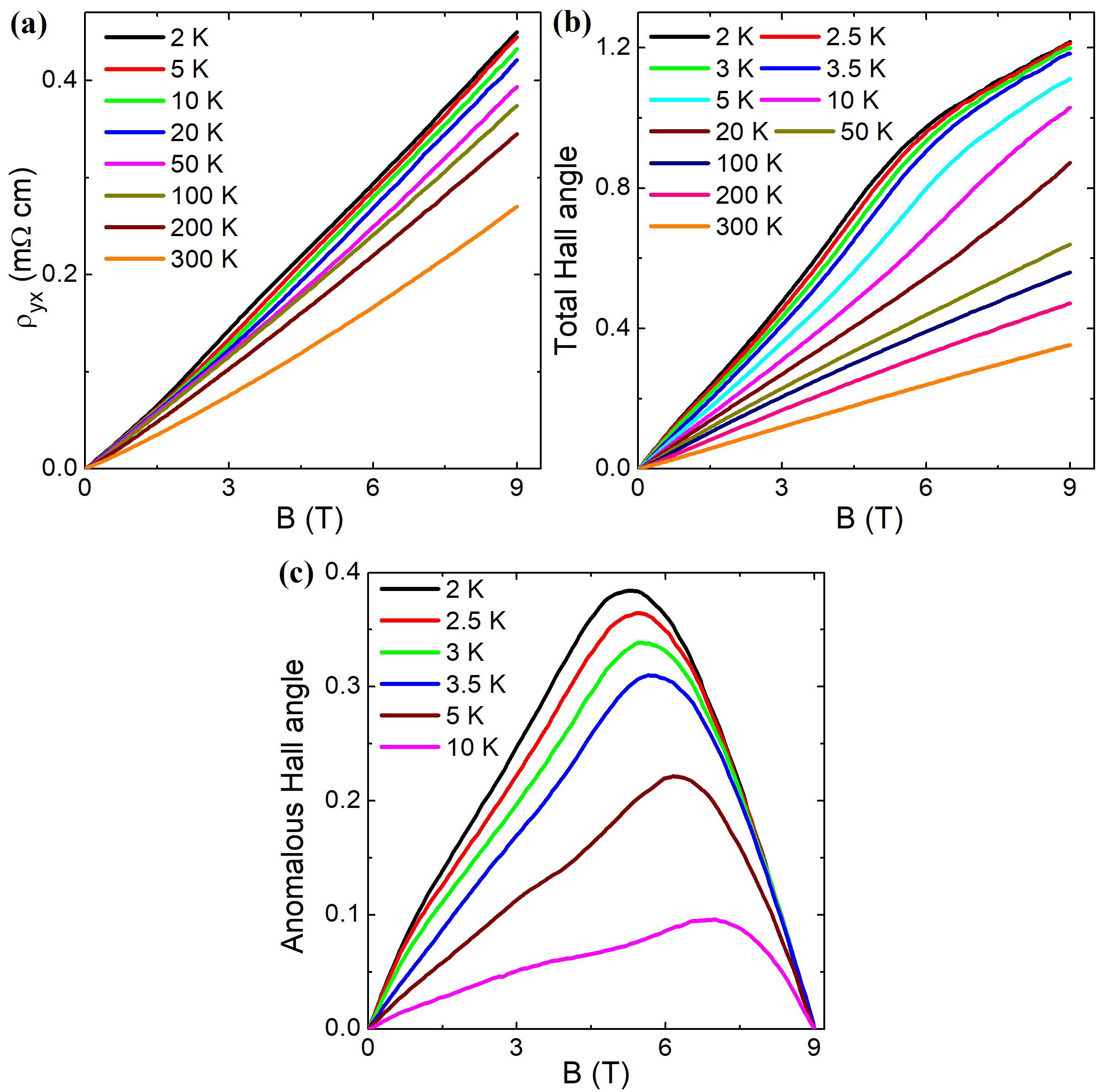}
\caption{(a) Magnetic field dependence of Hall resistivity in TbPtBi. (b) The calculated total Hall angle at different temperatures. (c) The magnetic field dependence of the anomalous Hall angle.}
\end{figure}

\begin{center}
\begin{table}
\caption{Anomalous Hall conductivity ($\sigma^{A}_{xy}$) and anomalous Hall angle (AHA) for different compounds.}
 \begin{tabular}{|c c c c|}
 \hline
 Compound & $\sigma^{A}_{xy}$ & AHA & Reference\\

  & $\Omega^{-1}cm^{-1}$ &  & \\ [0.5ex]
 \hline\hline

 TbPtBi & 744 & 0.38 & This work\\
 \hline
 GdPtBi & 30-200 & 0.16-0.32 & \cite{Suzuki}\\
 \hline
 Co$_{3}$Sn$_{2}$S$_{2}$ & 1130 & 0.2 & \cite{Liu2}\\
 \hline
 Fe thin film & 1100 & 0.01 & \cite{Miyasato}\\
 \hline
 Ni thin film & 800 & 0.008 & \cite{Miyasato}\\
 \hline
 Gd thin film & 1000 & 0.02 & \cite{Miyasato}\\
 \hline
 Mn$_{3}$Ge & 500 & 0.05 & \cite{Nayak}\\
 \hline
 Mn$_{3}$Sn & 100 & 0.04 & \cite{Nakatsuji}\\
 \hline
 MnSi & 150 & 0.037 & \cite{Manyala}\\
 \hline
 MnGe & 400 & 0.012 & \cite{Kanazawa}\\
 \hline
 $L$1$_{0}$ FePt thin film & 1243 & 0.033 & \cite{Yu}\\
 \hline
\end{tabular}
\end{table}
\end{center}

\section{Conclusions}
To summarize, we have presented a systematic analysis on the magnetotransport properties of half-Heusler compound TbPtBi. This material hosts a frustrated antiferromagnetic ground state below 3.4 K. Large, non-saturating MR has been observed with unconventional magnetic field dependence. The MR shows significant anisotropy with crystallographic direction and hence, indicates the anisotropic nature of the Fermi surface. From the longitudinal MR, we have observed the signature of relativistic Adler-Bell-Jackiw chiral anomaly, which confirms the presence of Weyl fermions in this material. Hall resistivity measurements reveal hole type carriers along with very high mobility. At low temperature, a large anomalous Hall conductivity has been observed, which is originated from the non-trivial Berry curvature of the electronic band structure. The calculated anomalous Hall angle is largest among the members of the half-Heusler $R$PtBi family as well as other compounds studied so far. Therefore, TbPtBi is an ideal system to investigate the complex interplay of magnetism and topologically non-trivial quantum states.

\end{document}